\documentclass[journal]{IEEEtran}

\usepackage{cite}

\usepackage{multirow}
\usepackage{fixmath}
\usepackage{amssymb,amsmath,bm}
\usepackage{cases}

\ifCLASSINFOpdf
   \usepackage[pdftex]{graphicx}
\else
   \usepackage[dvips]{graphicx}
\fi
\ifCLASSOPTIONcompsoc
  \usepackage[caption=false,font=normalsize,labelfont=sf,textfont=sf]{subfig}
\else
  \usepackage[caption=false,font=footnotesize]{subfig}
\fi
\begin{document}
%
\title{Divide and Conquer: A Deep CASA Approach to Talker-independent Monaural Speaker Separation}
%
%
%

\author{Yuzhou~Liu 
        and~DeLiang~Wang
\thanks{This work was supported in part by two NIDCD Grants (R01 DC012048 and R01 DC015521), and in part by the Ohio Supercomputer Center.}
\thanks{Y. Liu is with the Department of Computer Science and Engineering, The Ohio State University, Columbus, OH 43210-1277 USA (e-mail: liuyuz@cse.ohio-state.edu).}
\thanks{
D. L. Wang is with the Department of Computer Science and Engineering and the Center for Cognitive and Brain Sciences, The Ohio State University, Columbus, OH 43210-1277 USA (e-mail: dwang@cse.ohio-state.edu).
}}

%
%

\markboth{}%
{Shell \MakeLowercase{\textit{et al.}}: Bare Demo of IEEEtran.cls for Journals}
%



\maketitle

\begin{abstract}

We address talker-independent monaural speaker separation from the perspectives of deep learning and computational auditory scene analysis (CASA).
Specifically, we decompose the multi-speaker separation task into the stages of simultaneous grouping and sequential grouping. 
Simultaneous grouping is first performed in each time frame by separating the spectra of different speakers with a permutation-invariantly trained neural network.
In the second stage, the frame-level separated spectra are sequentially grouped to different speakers by a clustering network.
The proposed deep CASA approach optimizes frame-level separation and speaker tracking in turn, and produces excellent results for both objectives.
Experimental results on the benchmark WSJ0-2mix database show that the new approach achieves the state-of-the-art results with a modest model size.

  \end{abstract}


%
\IEEEpeerreviewmaketitle

\section{Introduction}


      \IEEEPARstart{S}{peech} usually occurs simultaneously with interference in real acoustic environments.
   Interference suppression is needed in a wide variety of speech applications, including automatic speech recognition, speaker identification, and hearing aids. 
   One particular kind of interference is the speech signal from competing speakers.
  Although human listeners excel at attending to a target speaker even without any spatial cues \cite{human-ss1}, speech separation remains a challenge for machines despite decades of research. 
   In this study, we address monaural (one microphone) speaker separation, mainly in the case of two concurrent speakers, which is also known as co-channel speech separation.

A traditional approach to monaural speech separation is computational auditory scene analysis (CASA) \cite{CASA}, which is inspired by human auditory scene analysis (ASA) mechanisms \cite{ASA}. CASA addresses speech separation in two main stages: simultaneous grouping and sequential grouping. 
With an acoustic mixture decomposed into a matrix of time-frequency (T-F) units, simultaneous grouping aggregates T-F units overlapping in time to short segments, each originating from the same source. 
In sequential grouping, segments are grouped across time into auditory streams, each corresponding to a distinct source.
For example, an unsupervised speaker separation method \cite{Ke-unsupervised} first generates T-F segments based on pitch and onset/offset analysis, and then uses clustering to sequentially group T-F segments into speakers.

Recently deep learning has been employed to address speaker separation.
The general idea is to train a deep neural network (DNN) to predict T-F masks or spectra of two speakers in a mixture \cite{Du} \cite{Huang} \cite{Zhang}.
There are usually two output layers in such a DNN, one for an individual speaker. 
These studies assume that the two speakers do not change between training and testing.
It has been shown that such talker-dependent training leads to significant intelligibility improvement for hearing impaired listeners \cite{Masood}.
However, talker-dependent training does not generalize to untrained speakers. Talker-independent speaker separation has to address the permutation problem\cite{DC} \cite{PIT}, i.e., how the output layers are tied to the underlying speakers.
The details of the permutation problem are introduced in Section~\ref{sec:3.1}.

Frame-level permutation invariant training (denoted by tPIT) \cite{PIT} tackles this problem by examining all possible label permutations within each frame during training, and uses the one with the lowest frame-level loss to train the separation network.
A locally optimized output-speaker pairing can thus be reached, which leads to excellent frame-level separation performance.
However, the correct speaker assignment in tPIT's output may swap frequently across frames.
In other words, the frame-level optimized outputs cannot be readily streamed into underlying speakers without reorganization. 
To address this issue, an utterance-level PIT (uPIT) algorithm \cite{PIT} is proposed to align each speaker to a fixed output layer throughout a whole training utterance.
Recent uPIT improvements include new network structure \cite{CBLDNN} \cite{grid} and new training objectives \cite{CBLDNN}.
TasNet \cite{Tasnet} extends uPIT to the waveform domain using a convolutional encoder-decoder structure.
FurcaNeXt \cite{Furca} integrates gated activations and ensemble learning into TasNet, and reports very high performance.

Deep clustering (DC) \cite{DC} looks at the permutation problem from a different perspective.
In DC, a recurrent neural network (RNN) with bi-directional long short-term memory (BLSTM) is trained to assign one embedding vector to each T-F unit of the mixture spectrogram.
The Frobenius norm between the affinity matrix of embedding vectors and the affinity matrix of the ideal speaker assignment (or the ideal binary mask) is used as the training objective.
DC avoids the permutation problem due to the permutation-invariant property of affinity matrices.
As training unfolds, embedding vectors of T-F units dominated by the same source are drawn closer together, and embeddings of those units dominated by different sources become farther apart.
Clustering these embedding vectors using the K-means algorithm assigns each T-F unit to one of the speakers in the mixture, which can be viewed as binary masking for speech separation.
In \cite{DAN}, a concept of attractors is introduced to DC to enable ratio masking and real-time processing.
Alternative training objectives, together with a chimera network which simultaneously estimates DC embeddings and uPIT outputs, are proposed in \cite{alter}.
In \cite{end2end}, iterative phase reconstruction is integrated into the chimera network to alleviate phase distortions.
In \cite{tri}, a phase prediction network is further added to \cite{end2end} to estimate the clean phase of each speaker source.

DC and PIT represent major approaches to talker-independent speaker separation. There are, however, limitations.
As indicated in \cite{PIT} \cite{me}, uPIT sacrifices frame-level performance for better assignments at the utterance level.
The speaker tracking mechanism in uPIT works poorly for same-gender mixtures.
On the other hand, DC is better at speaker tracking, but its frame-level separation is suboptimal compared to ratio masking used in tPIT.

Inspired by CASA, PIT and DC, we proposed a deep learning based two-stage method in our preliminary study \cite{me} to perform talker-independent speaker separation.
The method consists of two stages, a simultaneous grouping stage and a sequential grouping stage.
In the first stage, a tPIT-BLSTM is trained to predict the spectra of the two speakers at each frame without speaker assignment.
This stage separates spectral components of the two speakers at the same frame, corresponding to simultaneous grouping in CASA.
In the sequential grouping stage, frame-level separated spectra and the mixture spectrogram are fed to another BLSTM to predict embedding vectors for the estimated spectra, such that the embedding vectors corresponding to the same speaker are close together, and those corresponding to different speakers are far apart.
A constrained K-means algorithm is then employed to group the two spectral estimates at the same frame across time to different speakers.
This stage corresponds to sequential grouping in CASA.

In this study, we adopt the same divide-and-conquer strategy but improve its realization in major ways, resulting in what we call a deep CASA approach. 
In the simultaneous grouping stage, we utilize a UNet \cite{UNET0} convolutional neural network (CNN) with densely-connected layers \cite{DENSE} to improve the performance of frame-level separation. 
A frequency mapping layer is added to deal with inconsistencies between different frequency bands. 
To overcome the effects of noisy phase in inverse short-time Fourier transform (STFT), we explore complex STFT objectives and time-domain objectives as the training targets.
In the sequential grouping stage, we introduce a new embedding representation and weighted objective function.
In addition, we leverage the latest development in temporal convolutional networks (TCNs) \cite{TCN1} \cite{TCN0} \cite{Tasnet} \cite{ashu}, and use a TCN for sequential grouping, which greatly improves speaker tracking.
A new dropout scheme is proposed for TCNs to overcome the overfitting problem.
The evaluation results and comparisons demonstrate the resulting system achieves better frame-level separation and speaker tracking at the same time  compared to uPIT and \cite{me}. 


    The rest of the paper is organized as follows. 
    Section ~\ref{sec:3} presents details on monaural speaker separation and permutation invariant training. 
    The proposed algorithm, including the simultaneous and sequential grouping stages, is introduced in Section~\ref{sec:4}.
    Section~\ref{sec:5} presents experimental results, comparisons and analysis. 
    Conclusion and related issues are discussed in Section ~\ref{sec:6}.

\section{Monaural speaker separation and permutation invariant training}
\label{sec:3}

\subsection{Monaural Speaker Separation}
\label{sec:3.1}

The goal of monaural speaker separation is to estimate \(C\) independent speech signals \(x_c(n)\), \(c = 1,..., C\), from a single-channel recording of speech mixture \(y(n)\), where \(y(n)=\sum_{c=1}^{C}{x_c(n)}\) and \(n\) indexes time. In this work, we focus on the co-channel situation where \(C=2\).

Many deep learning based speaker separation systems \cite{Du} \cite{Huang} \cite{Zhang} address this problem in the T-F domain, where STFT is calculated using an analysis window \(w(n)\) with FFT length \(N\) and frame shift \(R\):
\begin{equation}
\resizebox{.75 \linewidth}{!} {
$Y(t,f)=\sum_{n=-\infty}^{\infty}w(n-tR)y(n)e^{-j2{\pi}fn/N}$}
\end{equation}
\begin{equation}
\resizebox{.75 \linewidth}{!} {
$X_c(t,f)=\sum_{n=-\infty}^{\infty}w(n-tR)x_c(n)e^{-j2{\pi}fn/N}$}
\end{equation}
where \(t\) and \(f\) denote the frame and frequency, respectively. 
The magnitude STFT of the mixture signal \(|Y(t,f)|\), together with other spectral features, are fed into a neural network to predict a T-F mask \({M}_c(t,f)\) for each speaker \(c\).
The masks are multiplied by the mixture to estimate the original sources:
\begin{equation}
|\tilde{X}_c(t,f)|={M}_c(t,f)\odot|Y(t,f)|
\end{equation}
Here \(\odot\) denotes element-wise multiplication, and \(|\tilde{X}_c(t,f)|\) denotes the estimated magnitude STFT of speaker \(c\).
An estimate of complex STFT \(\hat{X}_c(t,f)\) can be obtained by coupling \(|\tilde{X}_c(t,f)|\) with noisy phase. 
In the end, separated waveforms are resynthesized using inverse STFT (iSTFT):
\begin{equation}
\resizebox{.91 \linewidth}{!} {
$\hat{x}_c(n)=\frac{\sum_{t=-\infty}^{\infty}w(n-tR)\frac{1}{N}\sum_{f=0}^{N-1}\hat{X}_c(t,f)e^{j2{\pi}fn/N}}{\sum_{t=-\infty}^{\infty}w^2(n-tR)}$}
\label{iSTFT}
\end{equation}

Various training targets of \(|\tilde{X}_c(t,f)|\) have been explored for masking based speech separation \cite{Targets}.
Phase-sensitive approximation (PSA) is found to be effective as it accounts for errors introduced by the noisy phase \cite{PSM} \cite{PIT}.
In PSA, the desired reconstructed signal is defined as: \(|{X}_c(t,f)|\odot{}cos(\phi_c(t,f))\), where \(\phi_c(t,f)\) is the phase difference between \(Y(t,f)\) and \(X_c(t,f)\).
Overall, the training loss at each frame is computed as:
\begin{equation}
    \label{eq:loss1}
    \resizebox{.91 \linewidth}{!} {
$J_t^{PSA}=\sum\limits_{f=1}^{F}\sum\limits_{c=1}^{2}||{M}_c(t,f)\odot|Y(t,f)|-|{X}_c(t,f)|\odot{}cos(\phi_c(t,f))||$}
\end{equation}
where \(||\cdot||\) denotes the \(l_1\) norm.

The above formulation works well only when each output layer is tied to a training target signal with similar characteristics.
For instance, we may tie each output to a specific speaker, leading to talker-dependent training. 
We may also tie two outputs with male and female speakers respectively, leading to gender-dependent training. 
However, for talker-independent training data, how to select output-speaker pairing becomes a nontrivial problem.
Think of a training set consisting of three female speakers.
For speaker 1-2 mixtures, we can tie output1 to speaker1, and output2 to speaker2.
For speaker 1-3 mixtures, again output1 can be tied to speaker1, and output2 tied to speaker3.
However, it is hard to decide the pairing arrangement for speaker 2-3 mixtures. 
If output-speaker pairing is not arranged properly, conflicting gradients may be generated during training, preventing the neural network from converging.
This is referred as the permutation problem \cite{DC} \cite{PIT}. 

\subsection{Permutation Invariant Training }
\label{sec:3.2}

Frame-level PIT \cite{PIT} overcomes the permutation problem by providing target speakers as a set instead of an ordered list,
and output-speaker pairing \(c\leftrightarrow\theta_c(t)\), for a given frame \(t\), is defined as the pairing that minimizes the loss function over all possible speaker permutations \(P\). 
For tPIT, the frame-level training loss in Eq.~\ref{eq:loss1} is rewritten as:
\begin{equation}
    \label{eq:loss2}
    \resizebox{.91 \linewidth}{!} {
$J_t^{tPIT-PSA}= \min\limits_{\theta_c(t)\in{P}}\sum\limits_{f,c} ||{M}_c\odot|Y|-|{X}_{\theta_c(t)}|\odot{}cos(\phi_{\theta_c(t)})||$}
\end{equation}
We omit \((t,f)\) in \({M}, Y, X,\) and \(\phi\) for brevity.

    
tPIT does a good job in separating two speakers at the frame level \cite{PIT} \cite{me}.
However, due to its locally optimized training objective, an output layer may be tied to different speakers at different frames, and the correct speaker assignment may swap frequently.
If we reassign the outputs with respect to the minimum loss for each speaker, tPIT can almost perfectly reconstruct both speakers \cite{me}.

Optimal speaker assignments are not obtainable in practice as the targets are not given beforehand. 
To address this issue, uPIT fixes output-speaker pairing \(c\leftrightarrow\theta_c(t)\) for a whole utterance,
which corresponds to the pairing that provides the minimum utterance-level loss over all possible permutations.

As reported in \cite{PIT} \cite{me}, uPIT considerably improves the separation performance with a default output assignment.
But it has the following shortcomings.
First, uPIT's output-speaker pairing is fixed throughout a whole utterance, which prevents frame-level loss to be optimized as in tPIT. As a result, uPIT always underperforms tPIT if their outputs are optimally reassigned. 
Second, uPIT addresses separation and speaker tracking simultaneously and due to limited modeling capacity of a neural network, uPIT does not work well for speaker tracking, especially for same-gender mixtures.

\section{Deep CASA approach to monaural speaker separation}
\label{sec:4}

We employ a divide and conquer idea to break down monaural speaker separation into two stages.
In the simultaneous grouping stage, a tPIT based neural network separates spectral components of different speakers at the frame-level.
The sequential grouping stage then streams frame-level estimates belonging to the same speaker.
Unlike uPIT, separation and tracking are optimized in turn in the deep CASA framework.
The two stages are detailed in the following subsections.

\subsection{Simultaneous Grouping Stage}
\label{sec:4.1}

\subsubsection{Baseline system}
\label{sec:4.1.1}
We adopt the tPIT framework described in \cite{me} as the baseline simultaneous grouping system.
The magnitude STFT of the mixture is used as the input.
BLSTM is employed as the learning machine.
The system is trained using the loss function in Eq.~\ref{eq:loss2}.
In the end, frame-level spectral estimates are passed to the second stage for sequential grouping.

\subsubsection{Alternative training targets for tPIT} 
\label{sec:4.1.2}

As mentioned, the PSA training target partially accounts for STFT phase, unlike the ideal binary mask (IBM) and ideal ratio mask (IRM).
However, PSA cannot completely restore the phase information in clean sources, because it uses noisy phase during iSTFT.
Recently, complex ratio masking \cite{CRM} (cRM) attempts to restore clean phase. 
The complex ideal ratio mask (cIRM) is defined in the complex STFT domain, with real and imaginary parts.
When applied to the complex STFT of the mixture, it perfectly reconstructs clean sources:
\begin{equation}
X_c(t,f)= \text{cIRM}_{c}(t,f) \otimes {Y(t,f)}
\end{equation}
where \(\otimes\) denotes point-wise complex multiplication. 

We propose complex ratio masking to perform monaural speaker separation.
Instead of directly using the cIRM as the training target, we first multiply the complex mixture by the estimated complex mask \(\text{cRM}_c\) to perform complex domain reconstruction:
\begin{equation}
\hat{X}_c(t,f)=\text{cRM}_c(t,f)\otimes Y(t,f)
\end{equation}
The reconstructed sources are then compared with clean sources to form the training objective:
\begin{equation}
    \label{eq:dsaf}
        \resizebox{.91 \linewidth}{!} {
$J_t^{tPIT-CA}= \min\limits_{\theta_c(t)\in{P}}\sum\limits_{f,c} \bold{[}  \hspace{0.01\linewidth} | \text{Re}(\hat{X}_c - {X}_{\theta_c(t)})|
+ |\text{Im}(\hat{X}_c - {X}_{\theta_c(t)})| \hspace{0.01\linewidth} \bold{]}$}
\end{equation}
where the \(l_1\) norm is applied to both the real and imaginary parts of the loss.
We call this training objective complex approximation (CA).

We also consider a training objective based on time-domain signal-to-noise ratio (SNR).
The proposed framework consists of two steps:
First, we organize all frame-level complex estimates \(\hat{X}_{c}\) with respect to the minimum frame-level loss, so that each organized output \(\hat{X}_{\theta_c(t)}\) corresponds to a single speaker. The frame-level loss for organization can be defined in three domains: the complex STFT, magnitude STFT and time domain. 
In each domain, we compare the estimates and ground-truth targets, and calculate the \(l_1\) norm of the difference as the loss.
We found the complex STFT loss to be slightly better.
Second, we apply iSTFT (Eq.~\ref{iSTFT}) to \(\hat{X}_{\theta_c(t)}(t,f)\), and compute utterance-level SNR for the final time-domain estimates \(\hat{x}_{\theta_c(t)}(n)\):
\begin{equation}
J^{tPIT-SNR}=\sum_{c=1}^2{10\hspace{0.01\linewidth} \text{log}\frac{\sum_n{x_c(n)^2}}{\sum_n{(x_c(n)-\hat{x}_{\theta_c(t)}(n))^2}}}
\end{equation}


\subsubsection{Convolutional neural networks for simultaneous grouping}
\label{sec:4.1.3}
    \begin{figure}[]
    \begin{minipage}[a]{\linewidth}
      \centering
      \centerline{\includegraphics[width=1.0\linewidth]{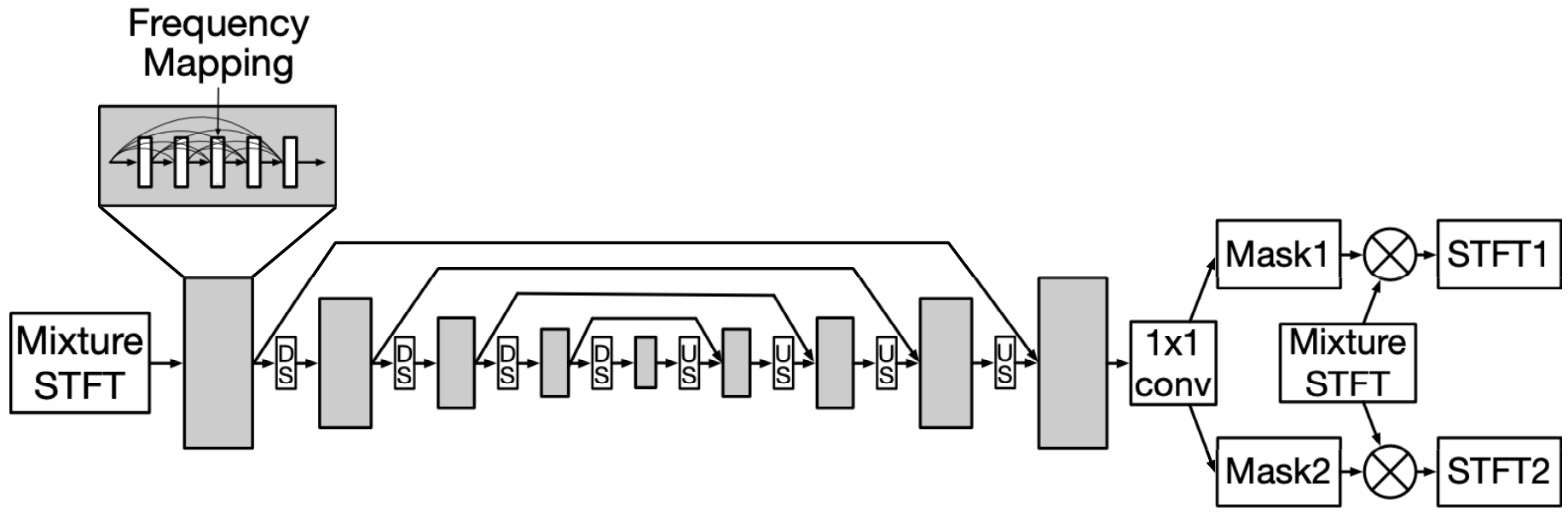}}
    \medskip
    \end{minipage}
    \caption{Diagram of the Dense-UNet used in simultaneous grouping. Gray blocks denote dense CNN layers. DS blocks denote downsampling layers and US blocks denote upsampling layers. Skip connections are added to connect layers at the same level. The inputs, masks and outputs can be defined in either magnitude or complex STFT domain.}
    \label{fig:unet}
    \end{figure}
    
Motivated by the recent success of DenseNet \cite{DENSE} and UNet \cite{UNET0} in music source separation \cite{UNET1} \cite{MM}, we propose a Dense-UNet structure for simultaneous grouping.

The proposed Dense-UNet is shown in Fig.~\ref{fig:unet}, and it is based on a UNet architecture \cite{UNET0}.
It consists of a series of convolutional layers, downsampling layers and upsampling layers.
The first half of the network encodes utterance-level STFT feature maps into a higher level of abstraction.
Convolutional layers and downsampling layers are alternated in this half, allowing the network to model large T-F contexts.
Convolutional layers and upsampling layers are alternated in the second half to project the encoded features back to its original resolution.
In this study, we use strided \(2\times 2\) depthwise convolutional layers \cite{depthwise} as downsampling layers. Strided transpose convolutional layers are used as upsampling layers. 
Skip connections are added between layers at the same hierarchical level in the encoder and decoder to preserve raw information in the mixture.

Next, we replace convolutional layers in the original UNet with densely-connected CNN blocks (DenseNet) \cite{DENSE}. 
The basic idea of DenseNet is to decompose one convolutional layer with many channels into a sequence of densely connected convolutional layers with fewer channels, where each layer is connected to every other layer in a feed-forward fashion:
\begin{equation}
z_l = H_{l} ([z_{l-1},z_{l-2},...,z_0])
\end{equation}
where \(z_0\) denotes the input feature map, \(z_l\) the output of the \(l^{th}\) layer, \([...]\) concatenation, and \(H_l\) the \(l^{th}\) convolutional layer followed by ELU (exponential linear unit) activation \cite{elu} and layer normalization \cite{ln}.
The DenseNet structure has shown excellent performance in image classification \cite{DENSE} and music source separation \cite{MM}.
In this study, all output layers \(z_l\) in a dense block have the same number of channels, denoted by \(K\). 
The total number of layers in each dense block is denoted by \(L\).
As shown in Fig.~\ref{fig:unet}, we alternate 9 dense blocks with 4 downsampling layers and 4 upsampling layers.
After the last dense block, we use a \(1 \times 1\) CNN layer to reorganize the feature map, and then output two masks.

In CNNs, convolutional kernels are usually applied across the entire input field. This is reasonable in the case of visual processing, where similar patterns can appear anywhere in the visual field with translation and rotation.
However, in the auditory representation of speech, patterns that occur in different frequency bands are usually different.
A generic CNN kernel may result in inconsistent outputs at different frequencies.
To address this problem, Takahashi and Mitsufuji \cite{MM} split the spectral input into several subbands, and train band-dependent CNNs, leading to a substantial rise in model size. 

We propose a frequency mapping layer which effectively alleviates this problem with a significant reduction of parameters.
The basic idea is to project inconsistent frequency outputs to an organized space using a fully-connected layer. 
We replace one CNN layer in each dense block with a frequency mapping layer.
The input to a frequency mapping layer is a concatenation of CNN layers \(z_l^0=[z_{l-1},z_{l-2},...,z_0]\in\mathbb{R}^{T\times F\times K'}\), where \(T\) and \(F\) denote time and frequency respectively, \(K'\) the number of channels in the input.
\(z_l^0\) is passed to a \(1\times 1\) convolutional layer, followed by ELU activation and layer normalization, to reduce the number of channels to \(K\). 
The resulting output is denoted by \(z_l^1 \in \mathbb{R}^{T\times F\times K}\). 
We then transpose the \(F\) and \(K\) dimension of \(z_l^1\) to get \(z_l^2 \in \mathbb{R}^{T\times K\times F}\).
Next, \(z_l^2\) is fed to a \(1\times 1\) convolutional layer, followed by ELU activation and layer normalization, to output \(z_l^3\in \mathbb{R}^{T\times K\times F}\). This layer can also be viewed as a frequency-wise fully connected layer, which takes all frequency estimates as the input and reorganize them in a different space.
Finally, \(z_l^3\) is transposed back, and the output of the frequency mapping layer \(z_l\in \mathbb{R}^{T\times F\times K}\) is generated.

\subsection{Sequential Grouping Stage}
\label{sec:4.2}
\subsubsection{Baseline system}
\label{sec:4.2.1}
In this stage, we group frame-level spectral estimates across time using a clustering network, which corresponds to sequential grouping in CASA.
In deep clustering based speaker separation, T-F level embedding vectors estimated by BLSTM are clustered into different speakers.
We extend this framework to frame-level speaker tracking.

\begin{figure}[]
    \begin{minipage}[a]{\linewidth}
      \centering
      \centerline{\includegraphics[width=0.5\linewidth]{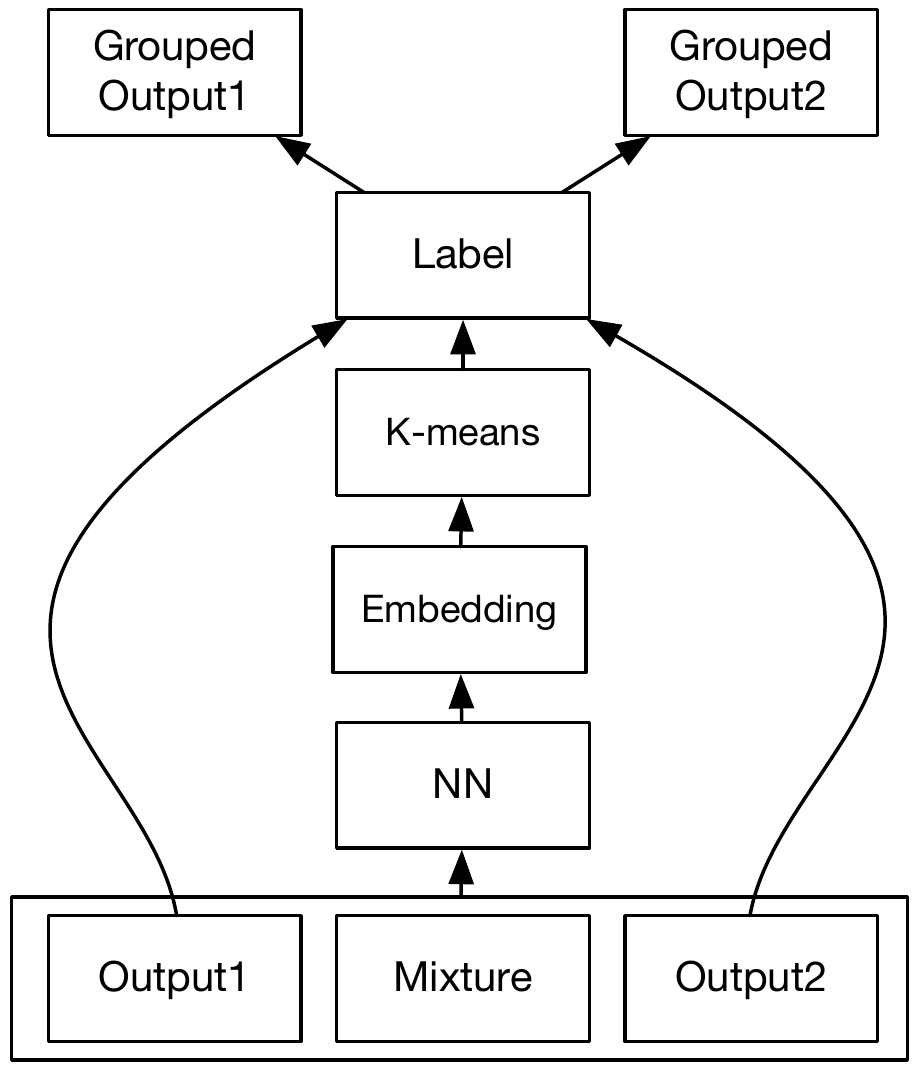}}
    \medskip
    \end{minipage}
    \caption{Diagram of the sequential grouping stage. We use BLSTM or TCN as the neural network in this stage.}
    \label{fig:sg}
    \end{figure}

Fig.~\ref{fig:sg} illustrates our sequential grouping. 
We first stack the mixture spectrogram and two spectral estimates (including real, imaginary and magnitude STFT) as the input to the system.
A neural network then projects frame-level inputs to a \(D\)-dimensional unit-length embedding vector \({\bold{V}}(t) \in \mathbb{R}^D\).
The target label is a two-dimensional indicator vector, denoted by \({\bold{A}}(t)\). 
During the training of tPIT, if the minimum loss is achieved when \(\hat{\bold{X}}_1(t)\) is paired with speaker 1, and \(\hat{\bold{X}}_2(t)\) is paired with speaker 2, we set \({\bold{A}}(t)\) to [1 0]. 
Otherwise, \({\bold{A}}(t)\) is set to [0 1].
In other words, \({\bold{A}}(t)\) indicates the optimal output assignment of each frame.
\({\bold{V}}(t)\) and \({\bold{A}}(t)\) can be reshaped into a \(T\times{D}\) matrix \(\bold{V}\), and a \(T\times{2}\) matrix \(\bold{A}\), respectively.
A permutation independent objective function \cite{DC} is:
\begin{equation}
J^{DC}=||\bold{V}\bold{V}^T-\bold{A}\bold{A}^T||^2_F
\end{equation}
where \(||\cdot||_F\) is the Frobenius norm.
Optimizing \(J^{DC}\) forces \({\bold{V}}(t)\) corresponding to the same optimal assignment to get closer during training, and otherwise to become farther apart.

Because we care more about the speaker assignment of frames where the two outputs are substantially different, a weight \(
w(t) = \frac{|LD(t)|}{\sum_t{|LD(t)|}}
\) is used during training
where \(LD(t)\) represents the frame-level loss difference (LD) between the two possible speaker assignments.
 \(LD(t)\) is large if two conditions are both satisfied:
 1) the frame-level energy of the mixture is high;
 2) the two frame-level outputs, \(\hat{X}_1(t,f)\) and \(\hat{X}_2(t,f)\), are quite different, so that the losses with respect to different speaker assignments are significantly different. 
\(w(t)\) can be used to construct a diagonal matrix \(\bold{W} = diag(w(t))\).
The final weighted objective function is:
\begin{equation}
J^{DC-W}=||\bold{W}^{1/2}(\bold{V}\bold{V}^T-\bold{A}\bold{A}^T)\bold{W}^{1/2}||^2_F
\label{eq:13}
\end{equation}
This objective function emphasizes frames where the speaker assignment plays an important role.

During inference, the K-means algorithm is first applied to cluster \({\bold{V}}(t)\) into two groups.
We then organize frame-level outputs according to their K-means labels. Finally, iSTFT is employed to convert complex outputs to the time domain.

    \begin{figure*}[]
    \begin{minipage}[a]{\linewidth}
      \centering
      \centerline{\includegraphics[width=0.95\linewidth]{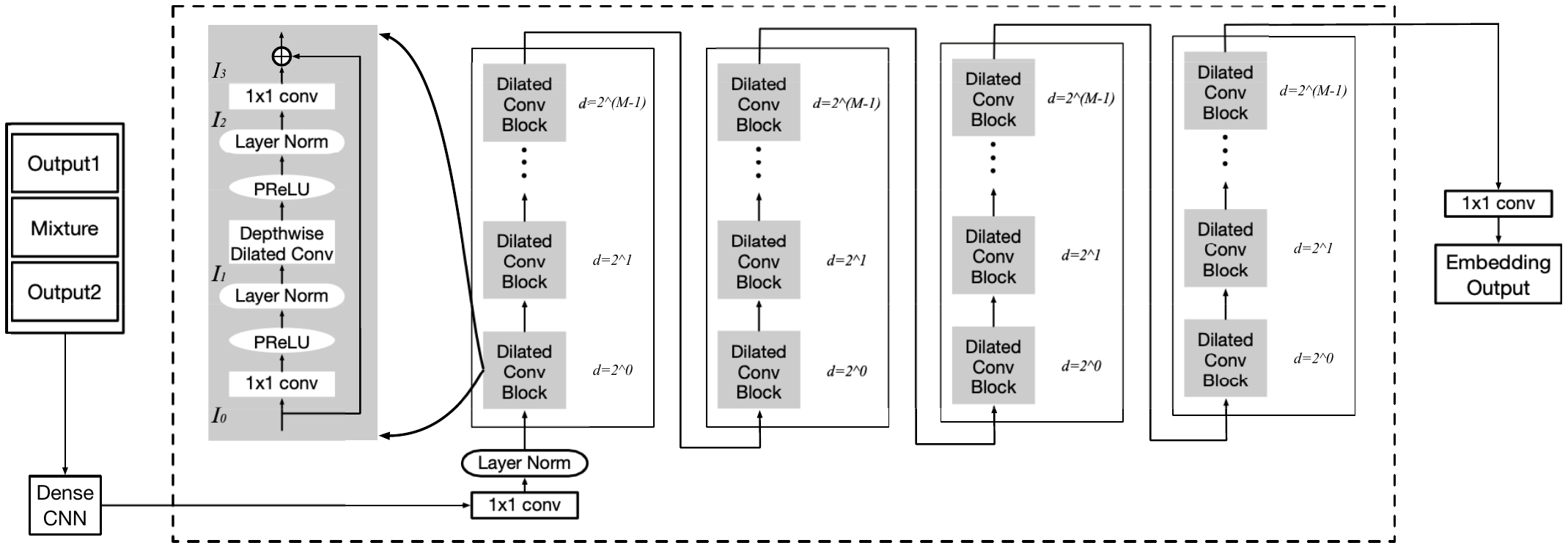}}
    \medskip
    \end{minipage}
    \caption{Diagram of the TCN used in sequential grouping. Outputs from the previous stage are fed into a series of dilated convolutional blocks to predict frame-level embedding vectors. The dilation factor of each block is marked on the right. The detailed structure of a dilated convolutional block is illustrated in the large gray box. The network within the dashed box can be also used for uPIT based speaker separation.}
    \label{fig:TCN}
    \end{figure*}

\subsubsection{Temporal convolutional networks for sequential grouping}
\label{sec:4.2.1}

Temporal convolutional networks (TCNs) have been used as a replacement for RNNs, and have shown comparable or better performance in various tasks \cite{TCN1} \cite{TCN0} \cite{Tasnet} \cite{ashu}.
In TCNs, a series of dilated convolutional layers are stacked to form a deep network, which enables very long memory.
In this study, we adopt a TCN similar to TasNet \cite{Tasnet} for sequential grouping, as illustrated in Fig.~\ref{fig:TCN}.

In the proposed TCN, input features are first passed to a 2-D dense CNN block, a \(1\times 1\) convolutional layer and a layer normalization module, to perform frame-level feature preprocessing.
The \(1\times 1\) convolutional layer here refers to a 1-D CNN layer with a kernel size of 1.
The preprocessed features are then passed to a series of dilated convolutional blocks, with an exponentially increasing dilation factor (\(2^0\), \(2^1\), ..., \(2^{M-1}\)) to exploit large temporal contexts.
Next, the \(M\) stacked dilated convolutional blocks are repeated 3 times to further increase the receptive field.
Lastly, the outputs are fed into a \(1\times 1\) convolutional layer for embedding estimation.

In each dilated convolutional block, a bottleneck input with \(B\) channels \(I_0 \in \mathbb{R}^{T\times B}\) is first passed to a \(1\times 1\) convolutional layer, followed by PReLU (parametric rectified linear unit) activation \cite{prelu} and layer normalization, to extend the number of channels to \(H\), with output denoted by \(I_1 \in \mathbb{R}^{T\times H}\).
A depthwise dilated convolutional layer \cite{depthwise} with kernel \(S \in \mathbb{R}^{3 \times H}\), followed by PReLU activation and layer normalization, is then employed to capture the temporal context.
The number 3 here indicates the size of the temporal filter in each channel, and there are \(H\) depthwise separable filters in the kernel.
We adopt non-causal filters to exploit both past and future information, with a dilation factor from \(2^0\),... \(2^{M-1}\), as in \cite{Tasnet}.
The output of this part is denoted by \(I_2 \in \mathbb{R}^{T\times H}\), which is then passed to a \(1\times 1\) convolutional layer to project the number of channels back to \(B\), denoted by \(I_3 \in \mathbb{R}^{T\times B}\).
In the end, an identity residual connection combines \(I_3\) and \(I_0\) and forms the final output. 

Overfitting is a major concern in sequence models.
If not regularized properly, sequence models tend to memorize the patterns in the training data, and get trapped in local minima. 
To address this issue, various dropout techniques \cite{RDO1} \cite{RDO3} \cite{RDO2} have been proposed for RNNs. 
Consistent improvement has been achieved if dropout is applied to recurrent connections \cite{RDO3}. 
Meanwhile, a simple dropout scheme for TCNs is used in \cite{TCN1}, i.e., dropping \(I_3\) in each dilated convolutional block, but it does not yield satisfactory performance in our experience. 
Based on these findings, we design a new dropout scheme for the TCN model, denoted by dropDilation.
In dropDilation, the dilated connections in depthwise dilated convolutional layers are dropped with a probability of \((1-p)\), where \(p\) denotes the keep rate.
To be more specific, a binary mask, \(
\bold{m} = [m_{-d} \quad 1 \quad m_{d}]^T \in \mathbb{R}^{3 \times 1} \), 
is multiplied with each depthwise dilated convolutional kernel \(S \in \mathbb{R}^{3 \times H}\) during training, with \(m_{-d} \) and \(m_{d} \) drawn independently from a Bernoulli distribution \(Bernoulli(p)\).
In dropDilation, we only drop the dilated connections while keeping the direct connections to preserve local information.

\section{Evaluation and comparison}
\label{sec:5}
\subsection{Experimental Setup}
\label{sec:5.1}
We use the WSJ0-2mix dataset, a monaural two-talker speaker separation dataset introduced in \cite{DC}, for evaluations.
WSJ0-2mix has a 30-hour training set and a 10-hour validation set generated by selecting random speaker pairs in the Wall Street Journal (WSJ0) training set si\_tr\_s, and mixing them at various SNRs between 0 dB and 5 dB.
Evaluation is conducted on the 5-hour open-condition (OC) test set, which is similarly generated using 16 untrained speakers from the WSJ0 development set si\_dt\_05 and si\_et\_05.
All mixtures are sampled at 8 kHz.
STFT with a frame length of 32ms, a frame shift of 8 ms, and a square root hanning window is taken for the whole system.

We report results in terms of signal-to-distortion ratio improvement (\(\mathrm{\Delta}\)SDR) \cite{SDR}, perceptual evaluation of speech quality (PESQ) \cite{PESQ}, and extended short-time objective intelligibility (ESTOI) \cite{eSTOI}, to measure source separation performance, speech quality and speech intelligibility, respectively.
We also report the final result in terms of scale-invariant signal-to-noise ratio improvement (\(\mathrm{\Delta}\)SI-SNR) \cite{Tasnet} for a systematical comparison with other competitive systems.

\subsection{Models}
\label{sec:5.2}
\subsubsection{Simultaneous grouping models}
\label{sec:5.2.1}
Two models are evaluated for simultaneous grouping: BLSTM and Dense-UNet.

The baseline BLSTM contains 3 BLSTM layers, with 896\(\times\)2 units in each layer. 
In each dense block of Dense-UNet, the number of channels \(K\) is set to 64, the total number of dense layers \(L\) is set to 5, and all CNN layers have a kernel size of \(3 \times 3\) and a stride of \(1\times 1\).
The middle layer in each dense block is replaced with a frequency mapping layer. 
We use valid padding (a term in CNN literature referring to no input padding) for the last CNN layer in each dense block, and same padding (padding the input with zeros so that the output has the same dimension as the original input) for all other layers.
The input STFT is zero-padded accordingly.

For both models, when trained with \(J_t^{tPIT-PSA}\), the magnitude STFT of the mixture is adopted as the input, and ELU activation is applied to output layers for phase-sensitive mask estimation.
If \(J_t^{tPIT-CA}\) or \(J^{tPIT-SNR}\) is used for training, a stack of real and imaginary STFT is used as the input, and linear output layers are used to predict the real and imaginary parts of complex ratio masks separately.

Both networks are trained with the Adam optimization algorithm \cite{Adam} and dropout regularization \cite{Dropout}.
The initial learning rate is set to 0.0002 for BLSTM, and 0.0001 for Dense-UNet.
Learning rate adjustment and early stopping are employed based on the loss on the validation set.

\subsubsection{Sequential grouping models}
\label{sec:5.2.2}
Two models are evaluated for sequential grouping: BLSTM and TCN.
Both models are trained on top of a well-tuned simultaneous grouping model.

The baseline BLSTM contains 4 BLSTM layers, with 300\(\times\)2 units in each layer. 
In TCN, the maximum dilation factor is set to \(2^6=64\), to reach a theoretical receptive field of 8.128s.
The number of bottleneck units \(B\) is selected as 256. 
The number of units in depthwise dilated convolutional layers \(H\) is set to 512.
Same padding is employed in all CNN layers.
DropDilation with \(p=0.7\) is applied during training.

A 2-D dense CNN block is used in both models for frame-level feature preprocessing, with \(K=16\), \(L=4\), a kernel size of \(1 \times 3\) \((T\times F)\) and a stride of \(1\times 1\).
The dimensionality of embedding vectors \(D\) is set to 40.
Both networks are trained with the Adam optimization algorithm, with an initial learning rate of 0.001 for BLSTM, and 0.00025 for TCN.
Learning rate adjustment and early stopping are again adopted.

\subsubsection{One stage uPIT models}
\label{sec:5.2.3}
To systematically evaluate the proposed methods, we train a Dense-UNet and a TCN with SNR objectives and uPIT training criterion, i.e., \(J^{uPIT-SNR}\).
Other training recipes follow those in Section~\ref{sec:5.2.1} and~\ref{sec:5.2.2}..

\subsection{Results and Comparisons}
\label{sec:5.3}

      \begin{table}[!t]
    \renewcommand{\arraystretch}{1.2}
    \caption{\(\mathrm{\Delta}\)SDR, PESQ and ESTOI for simultaneous grouping models with optimal output assignment on WSJ0-2mix OC.}
    \label{TABLE1}
    \centering
    \resizebox{\columnwidth}{!}{%
    {
    \begin{tabular}{l|c|c|c|c|c}
    \hline
      &Objective & \# of param. & \(\mathrm{\Delta}\)SDR (dB) & PESQ& ESTOI (\%) \\
    \hline
    Mixture & - &-& 0.0 & 2.02 & 56.1\\
   \hline
    tPIT BLSTM& PSA& 46.3M&13.0&3.13&86.7\\
    tPIT Dense-UNet &PSA & 4.7M&14.7	&3.41&90.5 \\
    tPIT Dense-UNet&CA & 4.7M& 18.6 & 3.57 & 93.8\\
    tPIT Dense-UNet&SNR& 4.7M& 19.1 & 3.63 & 94.3\\
    \hline

    \end{tabular}
    }
    }

    \end{table}

We first evaluate the simultaneous grouping stage.
Table~\ref{TABLE1} summarizes the performance of tPIT models with respect to different network structures and training objectives.
For all models, outputs are organized with the optimal speaker assignment before evaluation.
Scores of mixtures are presented in the first row. 
Compared to BLSTM, Dense-UNet drastically reduces the number of trainable parameters to 4.7 million, and introduces significant performance gain.
The frequency mapping layers in our Dense-UNet introduce a 0.3 dB increment in \(\mathrm{\Delta}\)SDR, 0.1 increment in PESQ, 0.8\% increment in ESTOI and a parameter reduction of 0.9 million. 
Next, we switch from magnitude STFT to complex STFT, and change the training objective to \(J_t^{tPIT-CA}\).
This change leads to large improvement, revealing the importance of phase information for source separation.
The SNR objective further outperforms the CA objective. 
We thus adopt tPIT Dense-UNet trained with \(J^{tPIT-SNR}\) for simultaneous grouping in the following evaluations.

 \begin{table}[!t]
    \renewcommand{\arraystretch}{1.2}
    \caption{\(\mathrm{\Delta}\)SDR, PESQ and ESTOI for tPIT and uPIT based Dense-UNet trained with SNR objectives.}
    \label{TABLE2}
    \centering
    \resizebox{\columnwidth}{!}{%
    {
    \begin{tabular}{l|c|c|c|c}
    \hline
     &Output Assign. & \(\mathrm{\Delta}\)SDR (dB) & PESQ& ESTOI (\%) \\
    \hline
    \multirow{2}{*}{tPIT Dense-UNet} & Optimal&19.1 & 3.63 & 94.3\\
     & Default & 0.0 & 1.99&55.8 \\
     \hline
     \multirow{2}{*}{uPIT Dense-UNet} &Optimal&17.0&3.40&91.6\\
     &Default&15.2 & 3.24 & 88.9       \\
    \hline 

    \end{tabular}
    }
    }

    \end{table}

\captionsetup[subfloat]{captionskip=-0.7pt}

\begin{figure*}[htb]
    \centering
    \subfloat[]{\includegraphics[width=0.5\linewidth]{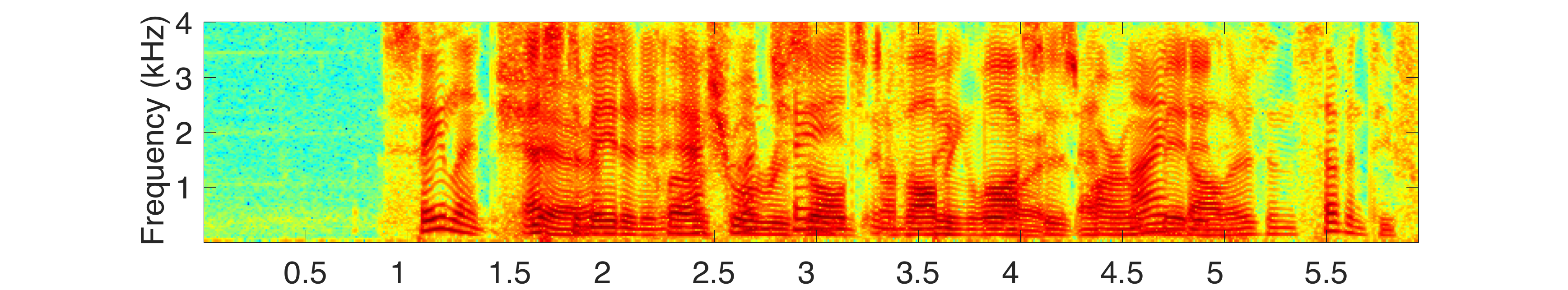}}
     \\[-2.2ex]
    \subfloat[]{\includegraphics[width=0.5\linewidth]{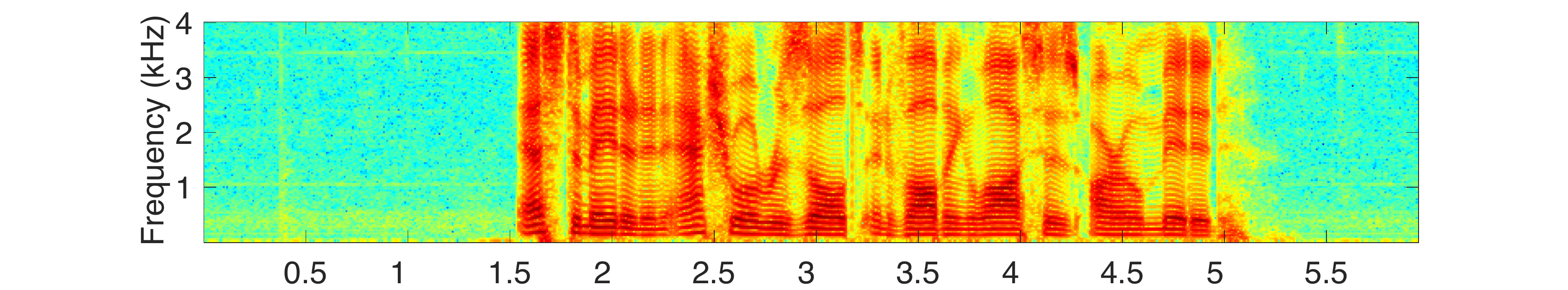}}
    \subfloat[]{\includegraphics[width=0.5\linewidth]{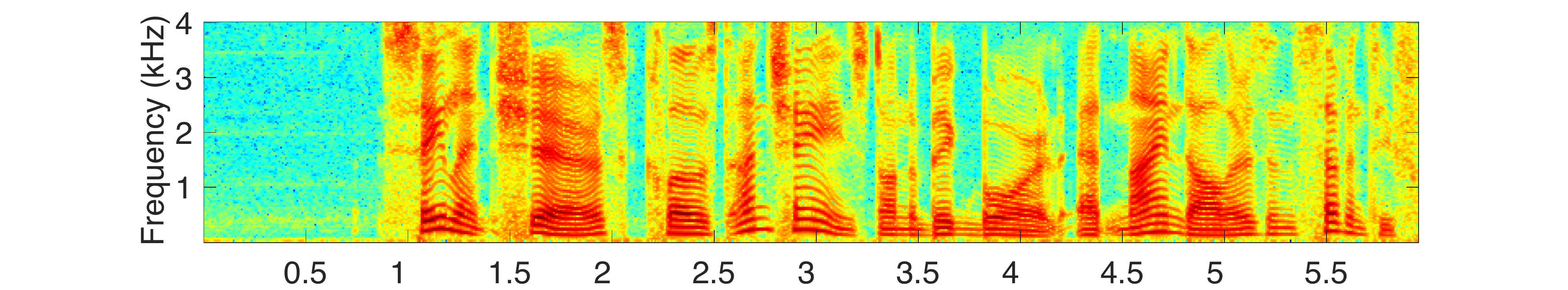}}
     \\[-2.2ex]
    \subfloat[]{\includegraphics[width=0.5\linewidth]{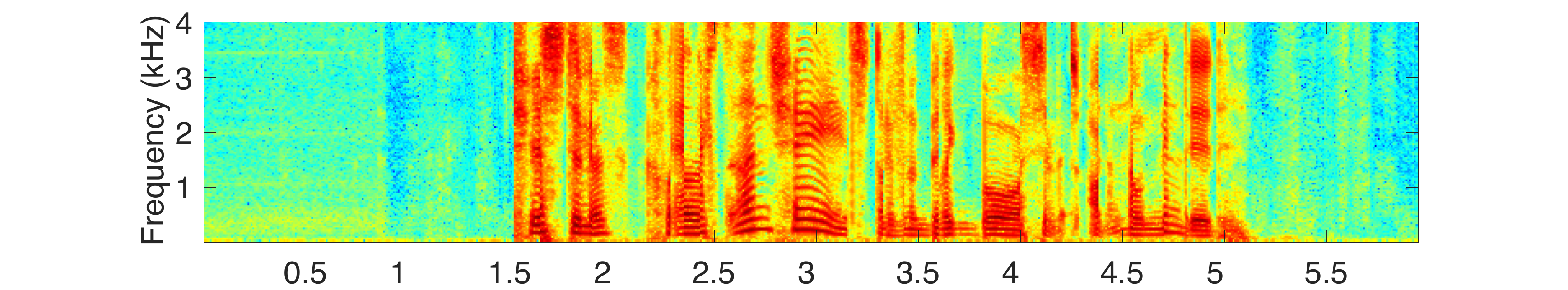}}
    \subfloat[]{\includegraphics[width=0.5\linewidth]{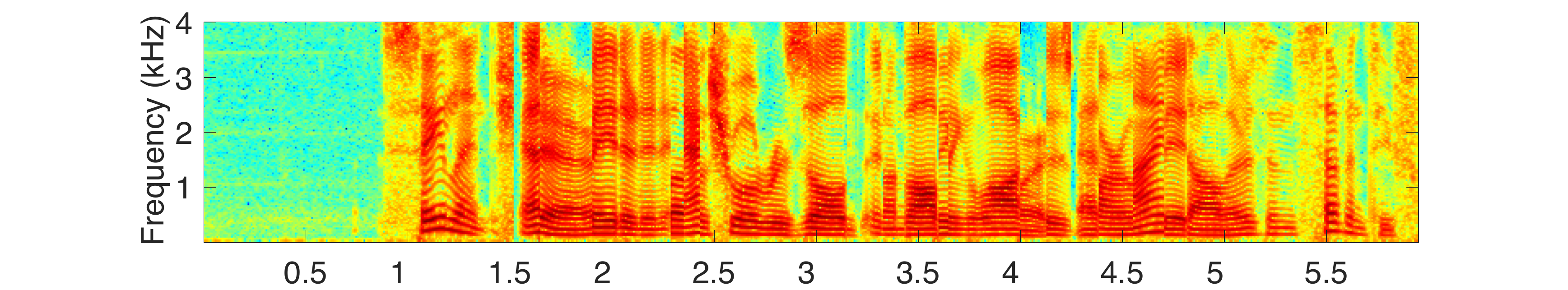}}
     \\[-2.2ex]
    \subfloat[]{\includegraphics[width=0.5\linewidth]{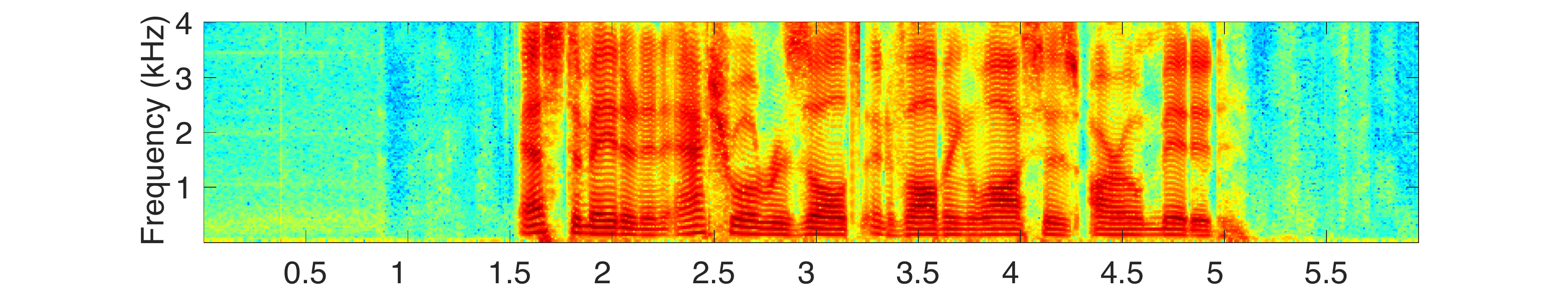}}
    \subfloat[]{\includegraphics[width=0.5\linewidth]{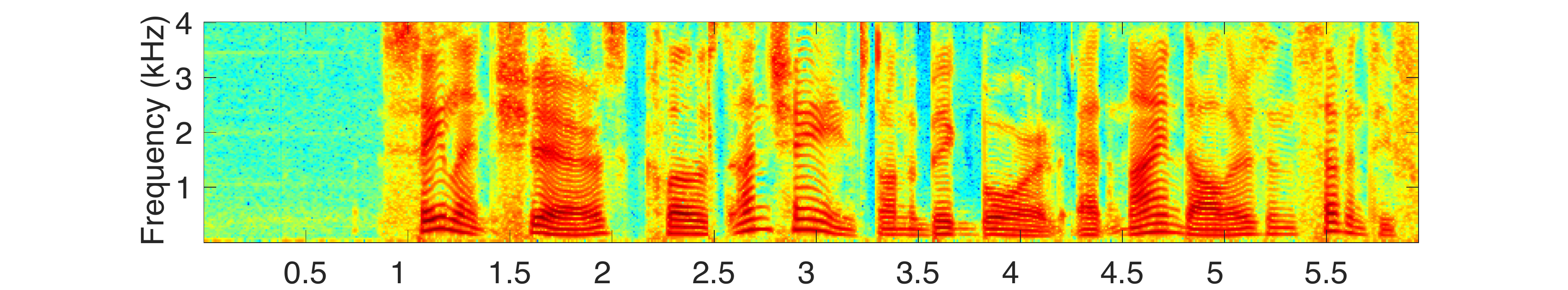}}
     \\[-2.2ex]
    \subfloat[]{\includegraphics[width=0.5\linewidth]{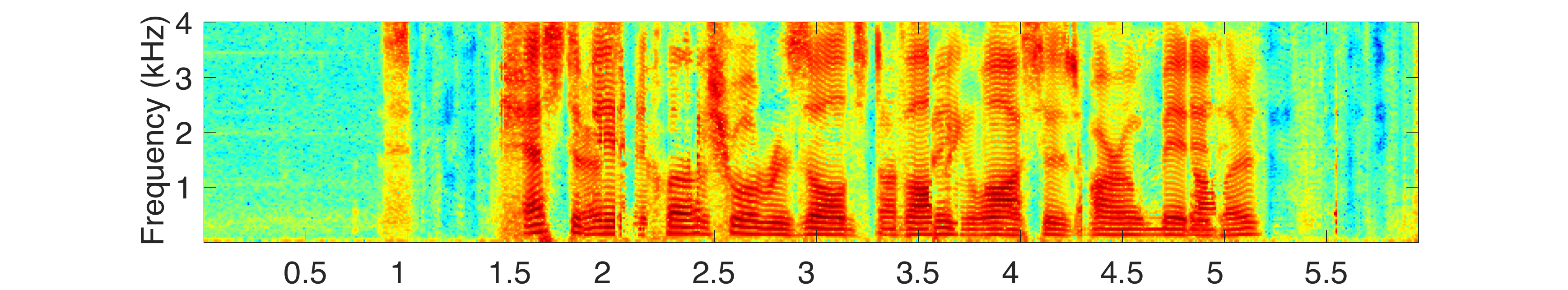}}
    \subfloat[]{\includegraphics[width=0.5\linewidth]{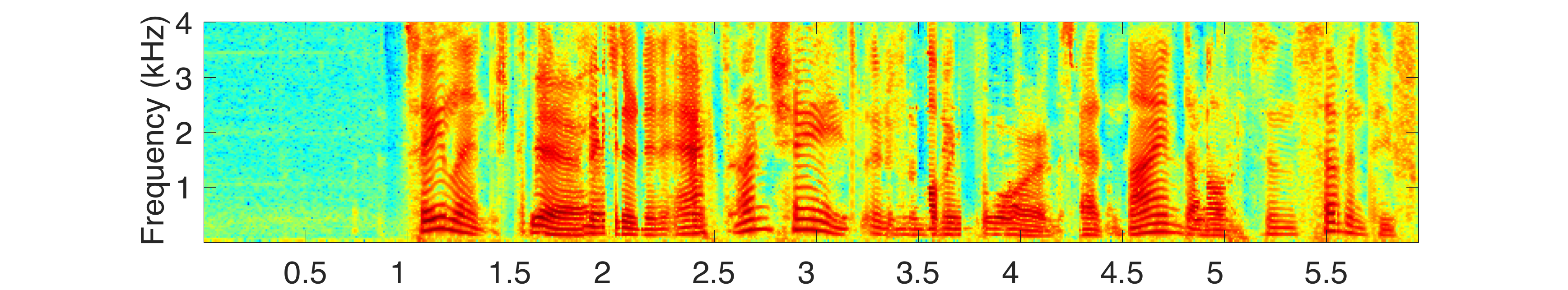}}
     \\[-2.2ex]
    \subfloat[]{\includegraphics[width=0.5\linewidth]{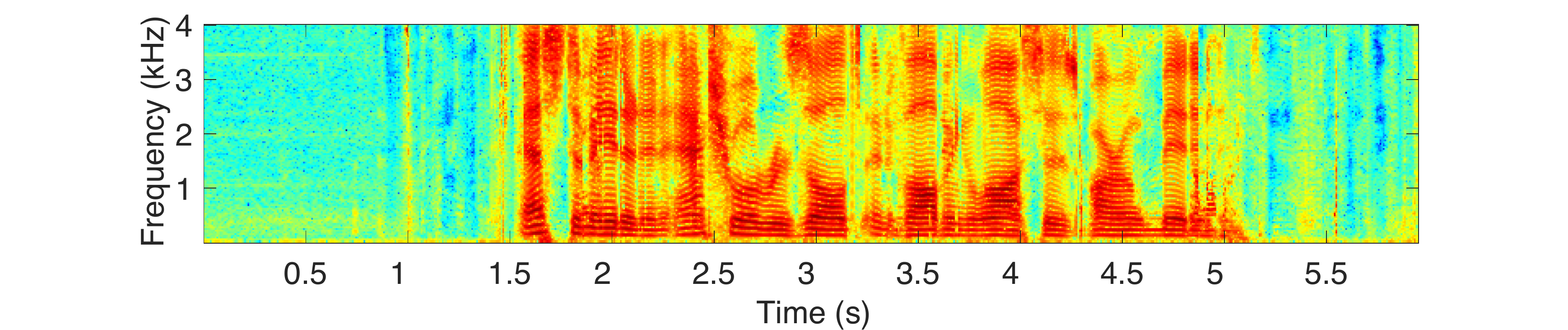}}
    \subfloat[]{\includegraphics[width=0.5\linewidth]{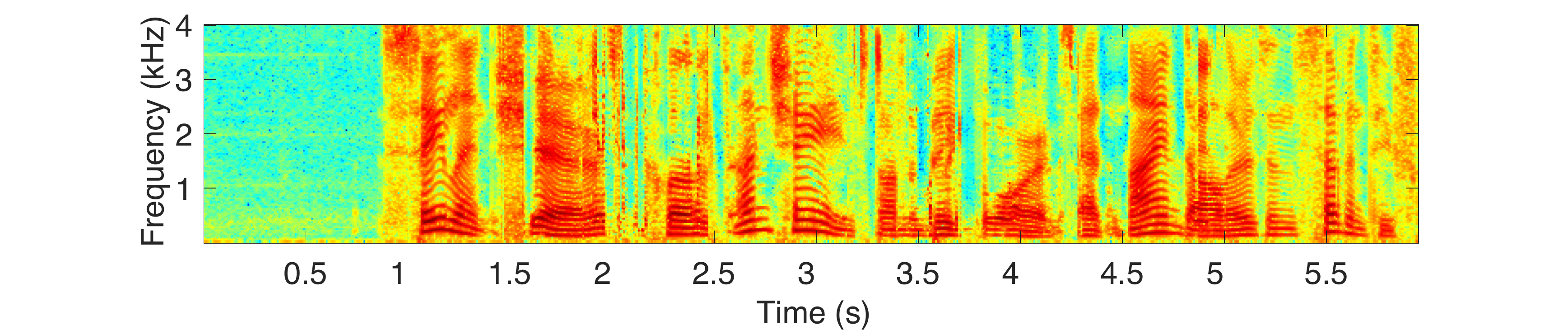}}
    \caption{Speaker separation results of PIT based models in log-scale magnitude STFT. Two models, tPIT Dense-UNet and uPIT Dense-UNet, are trained with CA objectives. The complex outputs from the models are converted to log magnitude STFT for visualization. (a) A male-male test mixture. (b) Speaker1 in the mixture. (c) Speaker2 in the mixture. (d) tPIT's output1 with default assignment. (e) tPIT's output2 with default assignment. (f) tPIT's output1 with optimal assignment. (g) tPIT's output2 with optimal assignment. (h) uPIT's output1 with default assignment. (i) uPIT's output2 with default assignment. (j) uPIT's output1 with optimal assignment. (k) uPIT's output2 with optimal assignment.}
    \label{fig:tpitresults}
 \end{figure*}

Table~\ref{TABLE2} compares tPIT and uPIT based Dense-UNet in terms of both optimal and default output assignments.
Both models are trained with SNR objectives.
Thanks to the utterance-level output-speaker pairing, uPIT's default assignment is improved by a large margin over tPIT.
However, since frame-level loss is not optimized in uPIT, there is a significant gap between uPIT and tPIT with optimal assignment.

Fig.~\ref{fig:tpitresults} illustrates the differences between tPIT and uPIT based Dense-UNet in more details.
Because SNR objectives lead to less structured outputs in the T-F domain, the models illustrated in the figure are trained with CA objectives.
Speaker assignment swaps frequently in the default outputs of tPIT. 
However, if we organize the outputs with the optimal assignment, the outputs almost perfectly match the clean sources, as shown in the fourth row. 
On the other hand, the default outputs of uPIT are much closer to the clean sources compared to tPIT.
However, for this same-gender mixture, uPIT makes several assignment mistakes in the default outputs, e.g., from 2s to 2.5s, and from 5s to 5.2s.
If we optimally organize uPIT's outputs, as in the last row, we can see uPIT exhibits much worse frame-level performance than tPIT.
In some frames, e.g., around 4.9s, the predicted frequency patterns are totally mixed up.
These observations reveal uPIT's limitations in both frame-level separation and speaker tracking for challenging speaker pairs.

      \begin{table}[!t]
   \renewcommand{\arraystretch}{1.2}
    \caption{Comparison of different sequential grouping methods on WSJ0-2mix OC.}
    \label{TABLE3}
    \centering
    \resizebox{\columnwidth}{!}{%
    {
    \begin{tabular}{l|c|c|c|c}
    \hline
     Simul. Group.&Seq. Group.& \(\mathrm{\Delta}\)SDR (dB) & PESQ& ESTOI (\%) \\
    \hline
     tPIT Dense-UNet& BLSTM&16.4 & 3.31 & 90.8\\
     tPIT Dense-UNet& TCN & 17.9 & 3.49&92.9 \\

     \hline
     uPIT Dense-UNet &-&15.2 & 3.24 & 88.9 \\
     uPIT Dense-UNet &Optimal&17.0&3.40&91.6 \\
     uPIT TCN &-& 13.5 & 3.06 & 85.9\\
     uPIT TCN &Optimal& 14.9 & 3.19 & 88.1\\
    \hline
    \end{tabular}
    }
    }
    \end{table}

Next, we evaluate different sequential grouping models in Table~\ref{TABLE3}. 
The first two models are trained on top of the tPIT Dense-UNet with the SNR objective. 
As shown in the table, TCN substantially outperforms BLSTM, both having around 8 million parameters. 
The dropDilation technique in our TCN introduces 0.5 dB \(\mathrm{\Delta}\)SDR gain compared to conventional dropout \cite{TCN1}.

In the last four rows of Table~\ref{TABLE3}, we report the results of uPIT models.
The first uPIT model is trained using Dense-UNet, and it significantly underperforms both deep CASA systems.
Even if the outputs are optimally reassigned, uPIT Dense-UNet still systematically underperforms deep CASA (tPIT Dense-UNet + TCN), due to its frame-level separation errors.
We also train a TCN model with uPIT objectives, and it yields much worse results than uPIT Dense-UNet.

To further analyze the differences between deep CASA and uPIT, we present frame assignment error (FAE) for the best performing deep CASA system and the two uPIT based models in Table~\ref{TABLE4}.
FAE is defined as the percentage of incorrectly assigned frames in terms of the minimum frame-level loss.
As shown in the table, uPIT Dense-UNet generates the highest FAE, because the network is not specifically designed for sequence modeling.
uPIT TCN slightly outperforms uPIT Dense-UNet due to its long receptive field. 
However, because uPIT TCN does not handle frequency patterns as well, its overall separation performance is worse than uPIT Dense-UNet.
Deep CASA cuts FAE by half compared to uPIT models. 
Such results demonstrate the benefits of the proposed divide-and-conquer strategy, which optimizes frame-level separation and speaker tracking in turn, and achieves better performance in both objectives.

         \begin{table}[!t]
    \renewcommand{\arraystretch}{1.2}
    \caption{Frame assignment errors for different methods for frames with significant energy (at least -20 dB relative to maximum frame-level energy).}
    \label{TABLE4}
    \centering
    \resizebox{0.8\columnwidth}{!}{%
    {
    \begin{tabular}{l|c|c}
    \hline
   Simul. Group. & Seq. Group. & Frame Assign. Errors (\%) \\
    \hline
     tPIT Dense-UNet & TCN &1.38\\
    uPIT Dense-UNet & -& 3.49 \\
     uPIT TCN&-& 3.07\\

    \hline
    \end{tabular}
    }
    }

    \end{table}
    
                  \begin{table}[!t]
    \renewcommand{\arraystretch}{1.2}
    \caption{\(\mathrm{\Delta}\)SDR, PESQ and ESTOI for deep CASA and uPIT for different gender combinations. }
    \label{TABLE5}
    \centering
    \resizebox{\columnwidth}{!}{%
    {
    \begin{tabular}{l|c|c|c|c}
    \hline
     Model &Gender Comb.& \(\mathrm{\Delta}\)SDR (dB) & PESQ& ESTOI (\%) \\

    \hline
     \multirow{3}{*}{ tPIT Dense-UNet + TCN Assign.} & Female-Male &18.9 & 3.57 & 93.9\\
    & Female-Female& {15.7} & {3.32}&{90.5} \\
    & Male-Male & {17.2} & {3.45}&{92.5} \\
      \hline
     \multirow{3}{*}{ uPIT Dense-UNet } & Female-Male &17.4 & 3.42 & 91.9\\
    & Female-Female& {10.9} & {2.89}& 83.5\\
    & Male-Male & {13.6} & {3.12}& 86.7\\
    \hline
     \multirow{3}{*}{ tPIT Dense-UNet + Opt. Assign.} & Female-Male &19.4 & 3.64 & 94.4\\
    & Female-Female& {18.8} & {3.61}&{93.9} \\
    & Male-Male & {18.7} & {3.62}&{94.3} \\
    \hline
    \end{tabular}
    }
    }

    \end{table}
    
          \begin{figure*}[htb]
    \centering
    \subfloat[]{\includegraphics[width=0.5\linewidth]{tpit_t1.pdf}}
    \subfloat[]{\includegraphics[width=0.5\linewidth]{tpit_t2.pdf}}
     \\[-2.2ex]
    \subfloat[]{\includegraphics[width=0.5\linewidth]{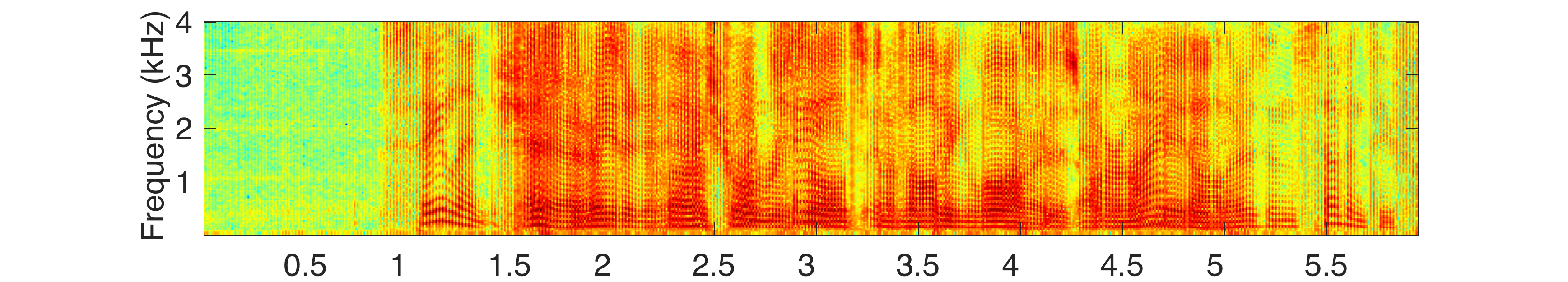}}
    \subfloat[]{\includegraphics[width=0.5\linewidth]{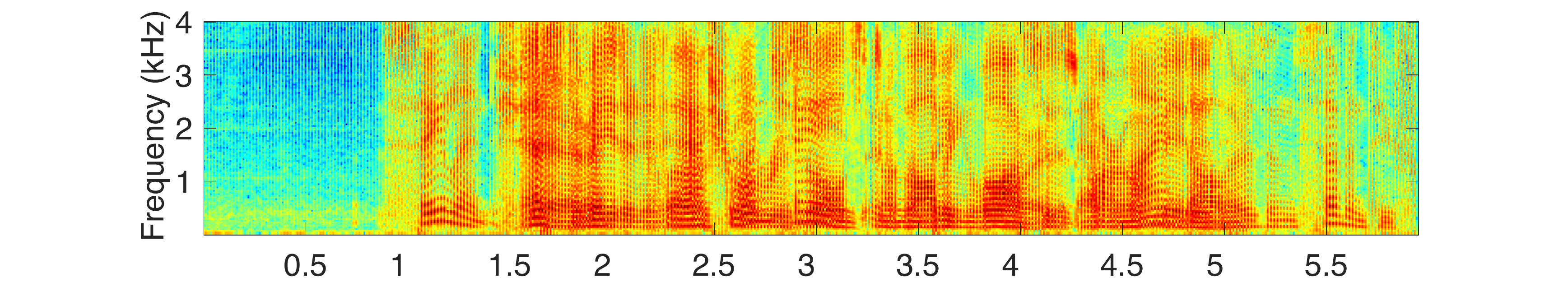}}
             \\[-0.1ex]

        \subfloat[]{\includegraphics[width=0.91\linewidth]{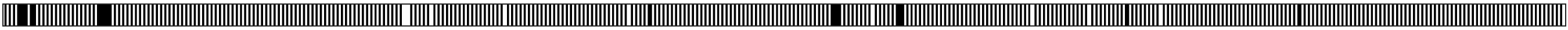}}
     \\[-0.2ex]

    \subfloat[]{\includegraphics[width=0.91\linewidth]{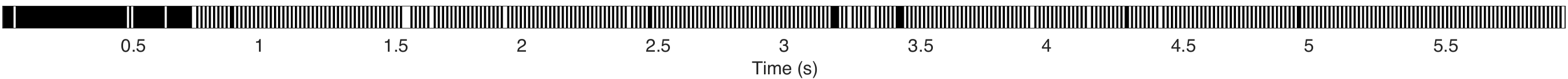}}
         \\[-2.5ex]

    \subfloat[]{\includegraphics[width=0.5\linewidth]{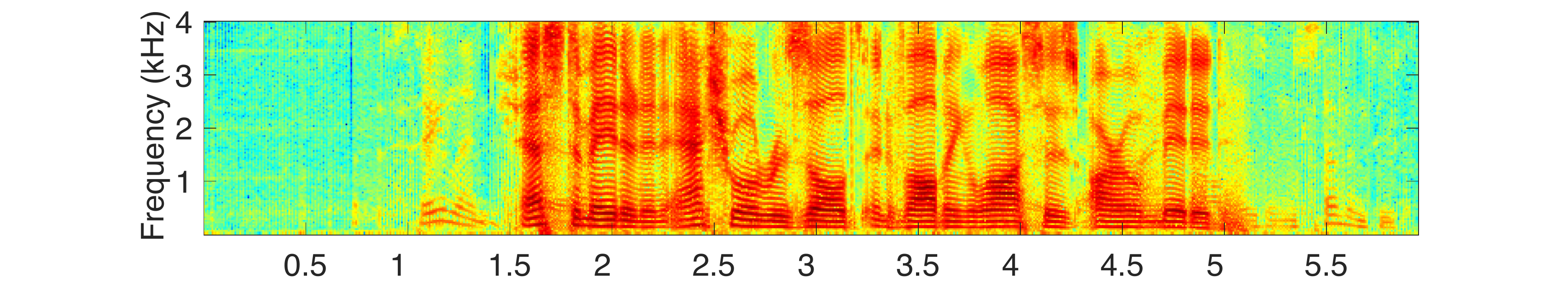}}
    \subfloat[]{\includegraphics[width=0.5\linewidth]{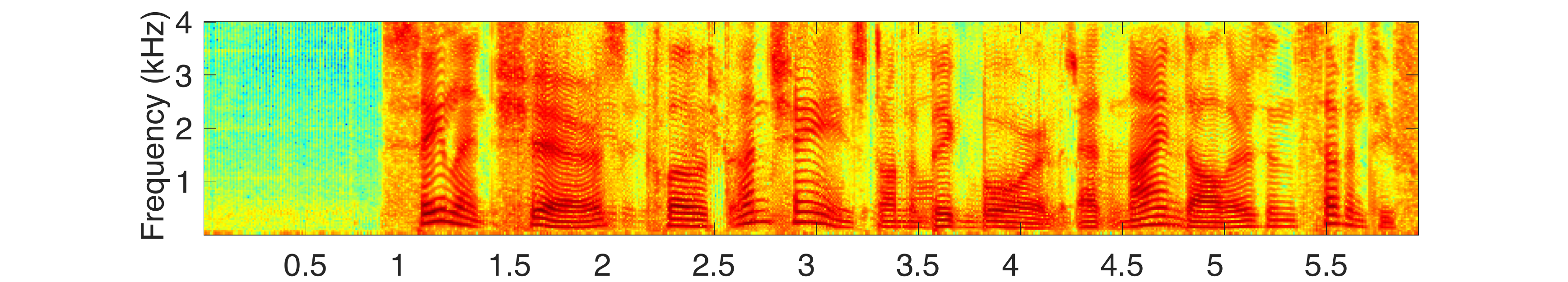}}
     \\[-2.2ex]
    \subfloat[]{\includegraphics[width=0.5\linewidth]{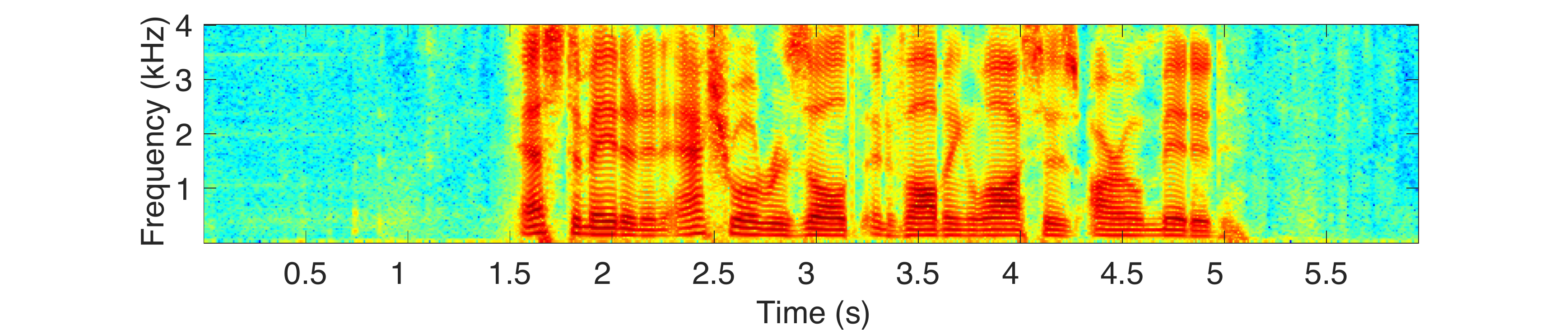}}
    \subfloat[]{\includegraphics[width=0.5\linewidth]{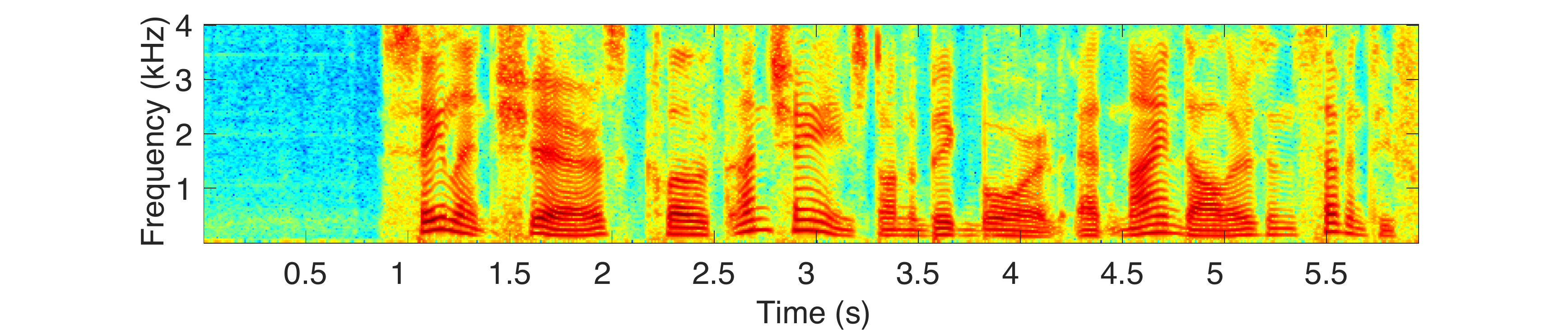}}
        \\[-0.1ex]
    
     \caption{Speaker separation results of the deep CASA system, with tPIT Dense-UNet trained with SNR objectives for simultaneous grouping and TCN for sequential grouping. The same test mixture is used as in Fig.~\ref{fig:tpitresults}. The complex outputs from the models are converted to log magnitude STFT for visualization. (a). Speaker1 in the mixture. (b) Speaker2 in the mixture. (c) tPIT's output1 with default assignment. (d) tPIT's output2 with default assignment. (e) Optimal assignment (black and white bars represent two different assignments). (f) K-means assignment. (g) tPIT's output1 with K-means assignment. (h) tPIT's output2 with K-means assignment. (i) tPIT's output1 with K-means assignment after iSTFT and STFT. (j) tPIT's output2 with K-means assignment after iSTFT and STFT.}
    \label{fig:CASA}

 \end{figure*}

          \begin{table*}[!t]
    \renewcommand{\arraystretch}{1.2}
    \caption{Number of parameters, \(\mathrm{\Delta}\)SDR, \(\mathrm{\Delta}\)SI-SNR, PESQ and ESTOI for various state-of-the-art systems evaluated on WSJ0-2mix OC. }
    \label{TABLE7}
    \centering
    \resizebox{1.5\columnwidth}{!}{%
    {
    \begin{tabular}{l|c|c|c|c|c}
    \hline
     & \# of param. & \(\mathrm{\Delta}\)SDR (dB) & \(\mathrm{\Delta}\)SI-SNR (dB)& PESQ& ESTOI (\%) \\
    \hline
    Mixture & - & 0.0 & 0.0 & 2.02 & 56.1\\
    uPIT \cite{PIT} & 94.6M&10.0 &-	&2.84&- \\
TasNet \cite{Tasnet}&8.8M&15.0&14.6&3.25&- \\
Wang et al. \cite{tri}&56.6M&15.4&15.2&3.45&-\\
FurcaNeXt \cite{Furca}&51.4M&\bf{18.4}&-&-&-\\
Deep CASA &12.8M&18.0&\bf{17.7}&\bf{3.51}&\bf{93.2}\\
    \hline
IBM&-&13.8&13.4&3.28&89.1\\
IRM &-&13.0&12.7&3.68&92.9\\
PSM&-&16.7&16.4&3.98&96.0\\
    \hline
    \end{tabular}
    }
    }    \end{table*}
    

Table~\ref{TABLE5} compares deep CASA and uPIT systems for different gender combinations. 
Both systems achieve better results on male-female combinations than same gender conditions.
The performance gap is larger for female-female mixtures, consistent with the observation in \cite{Listen}.
This might be due to the unbalanced gender distribution in WSJ0-2mix OC, which contains 1086 male-male mixtures, but only 394 female-female mixtures.
On the other hand, the performance gap between different gender combinations is much smaller in deep CASA than in uPIT, likely because deep CASA is better at speaker tracking.

Fig.~\ref{fig:CASA} illustrates the results of deep CASA.
As shown in the second row, tPIT Dense-UNet trained with SNR objectives generates entirely different default outputs compared to the same model trained with CA (cf. Fig.~\ref{fig:tpitresults}).
The optimal assignments alternate almost every frame, leading to striped patterns. 
To study the phenomenon, we analyze the overall training process of tPIT Dense-UNet trained with \(J^{tPIT-SNR}\).
At the beginning, the SNR objective leads to similar outputs as the CA objective.
However, because there is 75\% overlap between neighboring frames in the proposed STFT, models trained with SNR only need to make accurate predictions every other frame, with frames in between left blank.
Such patterns start to occur after a few hundred training steps.
The competing speaker then gradually fills in the blanks, and the striped patterns are thus formed.
As shown in Fig.~\ref{fig:CASA}(f), the K-means labels predicted by the sequential grouping system almost perfectly match the optimal labels in speech-dominant frames.
However, organizing the default outputs with respect to the K-means labels leads to magnitude STFT that is quite different from the clean sources.
Residual patterns from the interfering speaker still exist in some frames.
If we convert the complex outputs in Fig.~\ref{fig:CASA}(g) and 5(h) to the time-domain, these residual patterns will be cancelled by the overlap-and-add operation in iSTFT due to their opposite phases.
In the last row, we apply iSTFT and STFT in turn to the organized complex outputs, and the new results can almost perfectly match the clean sources.

 Simultaneous and sequential grouping are optimized in turn in the above deep CASA systems. 
We now consider joint optimization, where the two stages are trained together with small learning rates (1/8 of the initial learning rates) for 40 epochs. 
For the simultaneous grouping module, we organize the outputs using estimated K-means labels, and compare them with the clean sources to form an SNR objective.
Meanwhile, the sequential grouping module is trained using the weighted objective in Eq.~\ref{eq:13}.
As joint training unfolds, we observe smoother outputs.
Joint optimization introduces slight but consistent improvement in all three metrics (0.1 dB \(\mathrm{\Delta}\)SDR, 0.02 PESQ, and 0.3\% ESTOI). 

Finally, Table~\ref{TABLE7} compares the deep CASA system with joint optimization and other state-of-the-art talker-independent methods on WSJ0-2mix OC.
For all methods, we list the best reported results, and leave unreported fields blank.
The numbers of parameters in different methods are estimated according to the papers. 
The uPIT system \cite{PIT} is the basis of this study.
TasNet \cite{Tasnet} extends uPIT to the waveform domain, where a TCN is utilized for separation.
We have also trained a similar uPIT TCN in this work.
However, due to the different domains of signal representation, our uPIT TCN yields slightly worse results than TasNet, which suggests that better performance may be achieved by extending the deep CASA framework to the time domain.
In \cite{tri}, a phase prediction network is trained on top of a DC network. It yields high PESQ.
FurcaNeXt \cite{Furca} produces very high \(\mathrm{\Delta}\)SDR.
The deep CASA system generates slightly lower \(\mathrm{\Delta}\)SDR results, but has much fewer parameters.
In addition, deep CASA yields the best results in terms of \(\mathrm{\Delta}\)SI-SNR, PESQ and ESTOI.
The last three rows present the results of the IBM, IRM and ideal phase-sensitive mask (PSM) with the STFT configuration in Section~\ref{sec:5.1}.
Deep CASA systematically outperforms the ideal masks in terms of SDR and SI-SNR.
However, there is still room for improvement in terms of PESQ.

\section{Concluding remarks}
\label{sec:6}

We have proposed a deep CASA approach to talker-independent monaural speaker separation. 
Simultaneous grouping is first conducted to separate two speakers at the frame level. 
Sequential grouping is then employed to stream separated frame-level spectra into two sources. 
The deep CASA algorithm optimizes frame-level separation and speaker tracking in turn in the two-stage framework, leading to much better performance than DC and PIT.
Our contributions also include novel techniques such as complex ratio masking, SNR objectives, Dense-UNet with frequency mapping layers and TCN with dropDilation. 
Experimental results on the benchmark WSJ0-2mix dataset show that the proposed algorithm produces the state-of-the-art results, with a modest model size.

A major difference between our sequential grouping stage and deep clustering is that embedding operates at the T-F unit level in DC, and at the frame level in deep CASA.
There are several advantages to our approach.
First, DC excels at speaker tracking due to clustering, but it is not better than ratio masking for frame-level separation.
Therefore, divide and conquer is a natural choice.
Second, deep CASA is more flexible. 
Almost all DC based algorithms are built on time-frequency processing. 
Our sequential grouping works on frame-level outputs, which can be produced by estimating magnitude STFT, complex masks, or even time-domain signals.
In addition, we reduce the computational complexity of clustering from \(O(FT)\) in DC to \(O(T)\) in deep CASA. 


Although the deep CASA algorithm is formulated for two speakers, it can be extended to three or more speakers. 
First, additional output layers can be added in the simultaneous grouping stage.
In sequential grouping, we can employ the setup in \cite{me} to predict one embedding vector for each frame-level spectral estimate.
A constrained K-means algorithm can then be used to assign each frame-level embedding to a different speaker.



\ifCLASSOPTIONcaptionsoff
  \newpage
\fi



%

\bibliographystyle{IEEEtranS}

\bibliography{REFERENCE}

%

%





\end{document}